\documentclass[showpacs,preprintnumbers,amsmath,amssymb]{revtex4}

\usepackage{graphicx}
\usepackage{dcolumn}
\usepackage{bm}
\input{epsf.tex}

\def\be{\begin{equation}}
\def\ee{\end{equation}}
\def\ba{\begin{array}}
\def\ea{\end{array}}
\def\bea{\begin{eqnarray}}
\def\eea{\end{eqnarray}}
\begin{document}

\markboth{}{{\it Temeprature-dependent Seeger's liquid drop energy for nuclei up to Z=118}}

\title{Temeprature-dependent Seeger's liquid drop energy for nuclei up to Z=118}

\author{BirBikram Singh}
\altaffiliation{{\bf{Present Address}}: Institute of Physics, Sachivalaya Marg, Bhubaneswar-751005, India.}
\email{birbikram.singh@gmail.com}
\author{Manoj K. Sharma} \email{msharma@thapar.edu}
\affiliation{School of Physics and Materials Science, Thapar University, Patiala-147004, India.}
\author{Raj Kumar}
\author{Manie Bansal} 
\author{Raj K. Gupta}\email{rajkgupta.chd@gmail.com} 
\affiliation{Department of Physics, Panjab University, Chandigarh 160014, India.}
\date{\today}

\begin{abstract}

Seeger's semi-empirical mass formula is revisited for two of its constants (bulk constant $\alpha(0)$ and neutron-proton
asymmetry constant $a_a$) readjusted to obtain the ground-state (g.s.) binding energies of nuclei within a precision of 
$<$1.5 MeV and for nuclei up to Z=118. The aim is to include the temperature T-dependence on experimental binding energies, 
and not to obtain the new parameter set of Seeger's liquid drop energy $V_{LDM}$. Our proceedure is to define the g.s. 
binding energy  $B=V_{LDM}+\delta U$, as per Strutinsky renormalization procedure, and using the empirical shell 
corrections $\delta U$ of Myers and Swiatecki, fit the constants of $V_{LDM}$ to obtain the experimental binding energy 
$B_{expt}$ or theoretically calculated $B_{theo}$ if data were not available. The T-dependence of the constants of 
$V_{LDM}$, is introduced as per the work of Davidson {\it et al.}, where the pairing energy $\delta(T)$ is modified as per 
new calculations on compound nucleus decays. The newly fitted constants of $V_{LDM}$ at T=0 are made available here for 
use of other workers interested in nuclear dynamics of hot and rotating nuclei. 

\end{abstract}


 
\maketitle


\section{Introduction}

Seeger's mass formula \cite{seeger} was given in 1961, with its constants fitted to ground-state (g.s) binding energies of
some 488 nuclei available at that time. The temperature T-dependence of these constants was later introduced by Davidson
{\it et al.} \cite{davidson} on the basis of thermodynamical considerations of the nucleus. These constants, however, need 
be fitted again since a large amount of data on experimental g.s. binding energies \cite{audi03}, and their theoretically 
calculated values \cite{moller95} for, not-yet observed, neutron- and proton-rich nuclei have now become available. 
Furthermore, the T-dependence of the constants, in particular the pairing constant $\delta (T)$, need be looked in to 
because of their recent un-successful use in calculating the decay properties of some excited compound nuclear systems 
\cite{rkgupta08}-\cite{bansal11}. Note that our aim here is not to obtain a new set of constants for Seeger's mass formula, 
but simply to include the T-dependence on experimental binding energies $B_{expt}$. For this purpose, a readjustment of 
only two of the four constants, the bulk constant $\alpha(0)$ and the neutron-proton asymmetry constant $a_a$, are enough 
to obtain the $B_{expt}$ within $<$1.5 MeV. A similar job was first done in \cite{balou03} for nuclei up to Z=56, and then 
in \cite{bir08} up to Z=97, but is redone here with an improved accuracy and up to Z=118. Thus, the domain of the work is 
extended to neutron-deficient and neutron-excess nuclides where $B_{expt}$ are not available, but theoretical binding 
energies $B_{theo}$ are available \cite{moller95}. These re-fitted constants have been successfully used in the number of 
recent calculations \cite{rkgupta08}-\cite{niyti10} for studying the decay of hot and rotating compound nucleus (CN) formed 
in heavy ion reactions over a wide range of incident centre-of-mass (c.m.) energies.                  

A brief outline of the Seeger's mass formula, and the methodology used to workout the temperature-dependent binding 
energies, are presented in section II. Possible applications of the liquid drop energy in heavy ion reaction studies are
also included in this section. The calculations and results are given in section III, together with the table of fitted 
constants, which could be of huge importance for people working in the relevant area of nuclear physics. Finally, the 
results are summarized in section IV.

\section{Temperature-dependent Seeger's mass formula and Applications}

According to the Strutinsky renormalization procedure, the binding energy $B$ of a nucleus at temperature T is the sum
of liquid drop energy $V_{LDM}(T)$ and shell corrections $\delta U(T)$ 
\begin{equation}
B(T)=V_{LDM}(T)+\delta U(T),
\label{eq:1}
\end{equation}
where $V_{LDM}$ is the semi-empirical mass formula of Seeger \cite{seeger}, with T-dependence introduced by Davidson 
{\it et al.} \cite{davidson}, and $\delta U$ taken as the ``empirical'' formula of Myers and Swiatecki \cite{myers}, also 
made T-dependent to vanish exponentially, 
\begin{equation}
\delta U(T)=\delta U \exp(-T^2/T_0^2),
\label{eq:2}
\end{equation}
with $T_0$=1.5 MeV \cite{jensen73}. Seeger's liquid drop energy $V_{LDM}$, with its T-dependence due to Davidson 
{\it et al.}, is
\begin{eqnarray}
V_{LDM}(A,Z,T)&=&\alpha(T)A+\beta(T)A^{\frac{2}{3}}+\Bigl(\gamma(T)-\frac{\eta(T)}{A^{\frac{1}{3}}}\Bigr)
\Bigl[\frac{I^{2}+2|I|}{A}\Bigr]{}\nonumber\\
&&+\Bigl(\frac{Z^{2}}{r_0(T)A^{\frac{1}{3}}}\Bigr)
\Bigl[1-\frac{0.7636}{Z^{\frac{2}{3}}}-\frac{2.29}{[r_0(T)A^{\frac{1}{3}}]^{2}}\Bigr]{}+\delta(T){\frac{f(Z,A)}{A^{\frac{3}{4}}}},
\label{eq:3}
\end{eqnarray}
with 
$$I=a_a(Z-N), \qquad a_a=1$$
and, respectively, for even-even, even-odd and odd-odd nuclei, 
$$f(Z,A)=(-1,0,1).$$ 
Seeger's constants of 1961 are \cite{seeger}: 
$$\alpha(0)=-16.11, \quad \beta(0)=20.21, \quad \gamma(0)=20.65, \quad \eta(0)=48.00 \quad \hbox{(all in MeV)}, $$
and the pairing energy $\delta(0)$=33.0 MeV from \cite{benedetti64}. In the following, the bulk constant $\alpha$(0), and 
the neutron-proton asymmetry constant $a_a$, are found enough to be readjusted/ refitted to obtain the $B_{expt}$.

\begin{figure}[ht]
\begin{center}
\vspace*{-0.3cm}
\includegraphics[width=0.62\columnwidth]{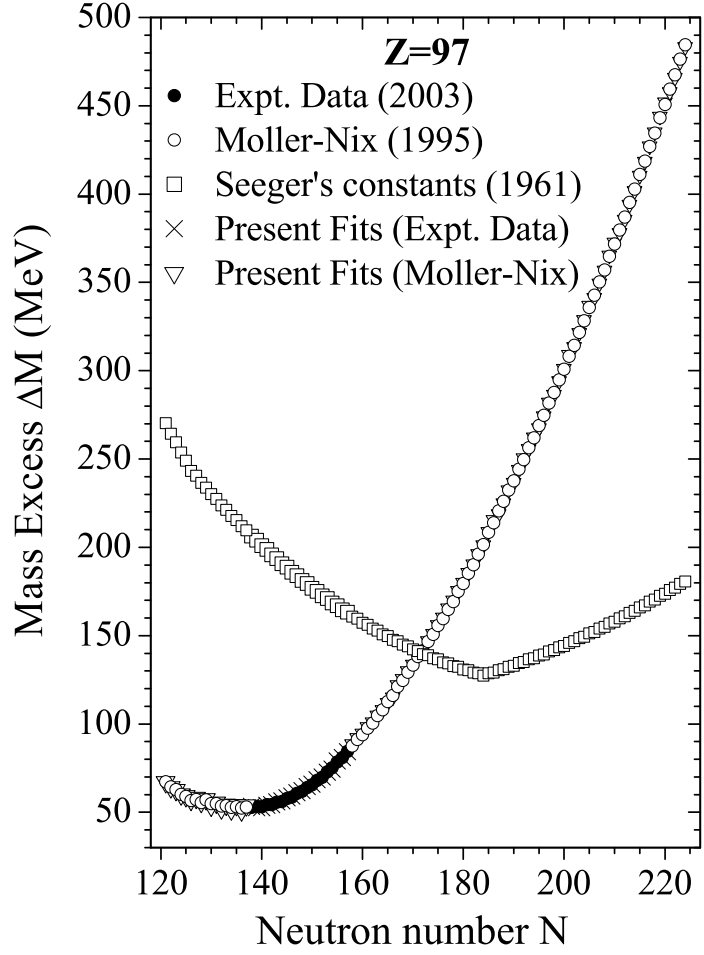}
\vspace*{-0.5cm}
\caption{Mass excess $\Delta M$ [=$M_A-A=NM_n+ZM_p+B(Z,N)-A$] in MeV as a function of neutron number N for Z=97, calculated 
by using the experimental data (solid circles) \cite{audi03}, theoretical data (open circles) \cite{moller95}, with newly 
fitted constants (crosses and down open triangles) and with the 1961 Seeger's constants \cite{seeger} (hollow squares).}
\label{fig:1}
\end{center}
\end{figure} 

\begin{figure}[ht]
\begin{center}
\vspace*{-1.0cm}
\includegraphics[width=0.6\columnwidth]{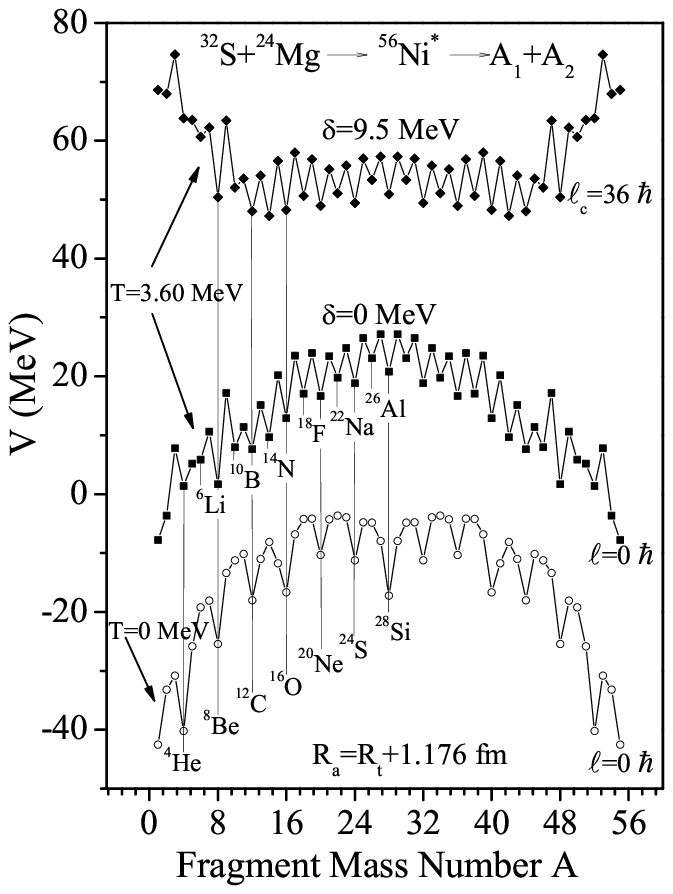}
\vspace*{-0.6cm}
\caption{Fragmentation potential [Eq. (\ref{eq:8})] calculated for the decay of $^{56}$Ni$^{*}$ formed in 
$^{32}$S+$^{24}$Mg reaction at T=3.60 MeV for $\ell$=0 and 36 $\hbar$, and also at T=0 for $\ell$=0 $\hbar$.}
\label{fig:2}
\end{center}
\end{figure} 

The T-dependence of the constants in Eq. (\ref{eq:3}) were obtained numerically by Davidson {\it et al.} \cite{davidson} 
from the available experimental information on excited states of 313 nuclei in the mass region 22$\le A\le$250 by 
determining the partition function ${\cal{Z}}(A,Z,T)$ of each nucleus in the canonical ensemble and making a least squares 
fit of the excitation energy
\begin{equation}
E_{ex}(A,Z,T)=V_{LDM}(A,Z,T)-V_{LDM}(A,Z,0)
\label{eq:4}
\end{equation}
to the ensemble average 
\begin{equation}
E_{ex}(A,Z,T)=T^2{\partial\over {\partial T}}ln{\cal{Z}}(A,Z,T).
\label{eq:5}
\end{equation}
The constants $\alpha (T)$, $\beta (T)$, $\gamma (T)$, $\eta (T)$ and $\delta (T)$ are given in Fig. 1 of Ref. 
\cite{davidson} for T$\le$4 MeV, extrapolated linearly for higher temperatures. However, $\delta (T)$ is constrained to be 
positive definite at all temperatures, and with $\delta (T)$=0 for T$>$2 MeV. Also, for the bulk constant $\alpha (T)$, 
instead, an empirically fitted expression using Fermi gas model is obtained, as
\begin{equation}
\alpha (T)=\alpha (0)+{T^2\over 15}.
\label{eq:6}
\end{equation}
 
For shell effects $\delta U$, the empirical formula of Myers and Swiatecki \cite{myers} is
\begin{equation}
\delta U=C\left [{{F(N)+F(Z)}\over {({\frac{A}{2}})^{2\over 3}}}-cA^{1\over 3}\right ] 
\label{eq:7}
\end{equation}
where
$$ F(X)={3\over 5} \left ( {{M_i^{5\over 3}-M_{i-1}^{5\over 3}} \over
{M_i-M_{i-1}}}\right) \left (X-M_{i-1}\right )-{3\over 5}\left
(X^{5\over 3}-M_{i-1}^{5\over 3}\right ) $$
with $X=N$ or $Z$, and $M_{i-1}<X<M_i$. $M_i$ are the magic numbers 2, 8, 14 (or 20), 28, 50, 82, 126 and 184 for both 
neutrons and protons. The constants $C$=5.8 MeV and $c$=0.26 MeV. Note that the above formula is for spherical shapes, but 
the missing deformation effects in $\delta U$ are included here to some extent via the readjusted constants of $V_{LDM}$ 
since we essentially use the experimental binding energies split in to two contributions, $V_{LDM}$ and $\delta U$, for 
reasons of adding the T-dependence on it.

Finally, as an application of the two components [$V_{LDM}(T)$ and $\delta U(T)$] of the (T-dependent) experimental binding 
energy in the field of heavy-ion reactons, we define the collective fragmentation potential 
\begin{eqnarray}
V(\eta,R,T)&=&\sum_{i =1}^{2}[V_{LDM}(A_i,Z_i,T)]+\sum_{i =1}^{2}[\delta U_i]\exp(-T^2/T^2_0){}+V_C(R,Z_i,\beta_{\lambda i},\theta_i,T)\nonumber\\
&&+V_P(R,A_i,\beta_{\lambda i},\theta_i,T){}+V_{\ell}(R,A_i,\beta_{\lambda i},\theta_i,T),
\label{eq:8}
\end{eqnarray}
where the nuclear proximity $V_P$, Coulomb $V_C$  and the angular-momentum $\ell$-dependent $V_{\ell}$ potentials are for 
deformed and oriented nuclei and are also T-dependent. For details, see, e.g., Ref. \cite{rkgupta08}.
 
Based on $V_{R,T}(\eta)$ at fixed R and T, and the scattering potential $V_{\eta,T}(R)$ at fixed $\eta$ and $T$, we 
calculate the CN decay cross-section by using the dynamical cluster-decay model (DCM) of Gupta and collaborators 
\cite{rkgupta08}-\cite{niyti10}, worked out in terms of the decoupled collective coordinates of mass (and charge) asymmetry 
$\eta={{A_1-A_2}\over {A_1+A_2}}$ [$\eta_Z={{Z_1-Z_2}\over {Z_1+Z_2}}$] and relative separation R. In terms of these 
coordinates, using $\ell$ partial waves, the CN decay cross section is defined as
\begin{equation}
\sigma ={\pi\over k^2}\sum _{\l =0}^{\l_{max}} (2l+1) P_0 P;\qquad k =\sqrt {2\mu E_{c.m.}\over \hbar^2} 
\label{eq:9}
\end{equation}
where the preformation probability $P_{0}$, refering to $\eta$ motion, is the solution of stationary Schr\"odinger equation 
in $\eta$ at a fixed R, and $P$, the WKB penetrability refers to R motion, both quantities carrying the effects of angular 
momentum $\ell$, temperature T, deformations $\beta _{\lambda i}$  and orientations $\theta _i$ degrees of freedom of 
colliding nuclei with c.m. energy $E_{c.m.}$. $\mu=[A_1A_2/(A_1+A_2)]m$, is the reduced mass, with m as the nucleon mass. 

Eq. (\ref{eq:9}) is applicable to the decay of CN to light particles (LPs, A$\le$4, Z$\le$2), intermediate mass fragments 
(IMFs, 2$\le$Z$\le$10), the fusion-fission fragments and the quasi-fission (q.f.) process where the incoming channel does 
not loose its identity, i.e., $P_0$=1 for qf. The $\ell_{max}$ could be fixed for the vanishing of the fusion barrier of 
the incoming channel, or the light particle cross-section $\sigma_{LPs}\rightarrow$0 , or else defined as the critical 
$\ell_{c}=R_a\sqrt {2\mu[E_{c.m.}-V(R_a,\eta_{in},\ell=0)]}/\hbar$. 

\begin{figure}[ht]
\begin{center}
\vspace*{-1.0cm}
\includegraphics[width=0.8\columnwidth]{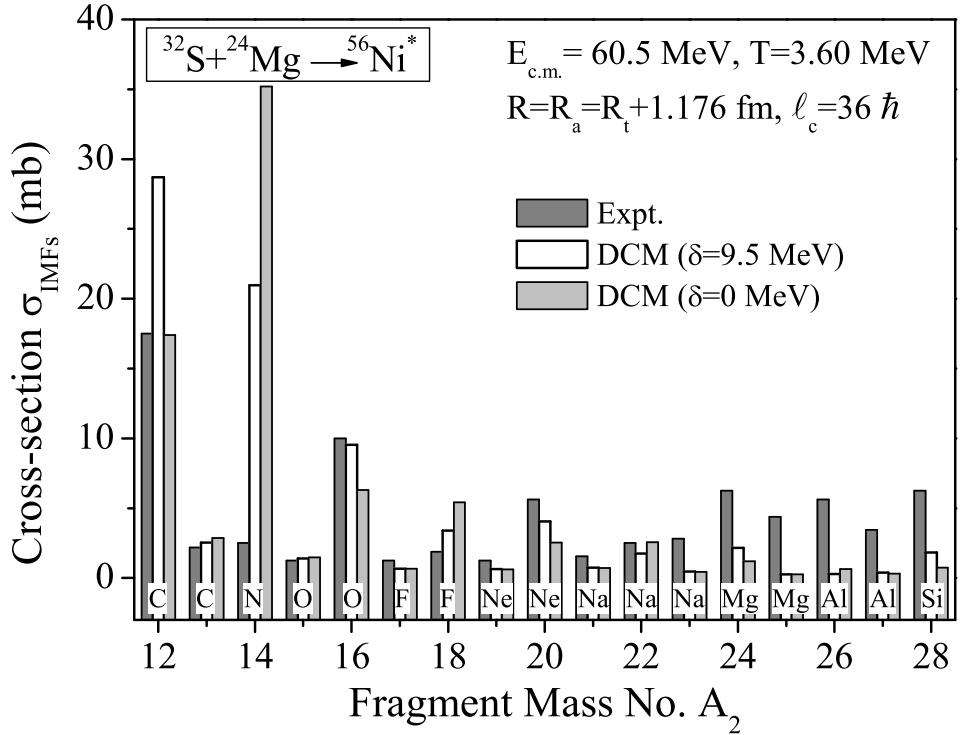}
\vspace*{-0.9cm}
\caption{The calculated IMFs cross-section $\sigma_{IMFs}$, using Eq. (\ref{eq:9}), for the decay of compound system 
$^{56}$Ni$^{*}$ formed in $^{32}$S+$^{24}$Mg reaction at T=3.60 MeV, taking pairing constant $\delta$=0 and 9.5 MeV,
compared with experimental data \cite{sanders89}.}
\label{fig:3}
\end{center}
\end{figure} 

\begin{figure}[ht]
\begin{center}
\vspace*{-0.7cm}
\includegraphics[width=0.6\columnwidth]{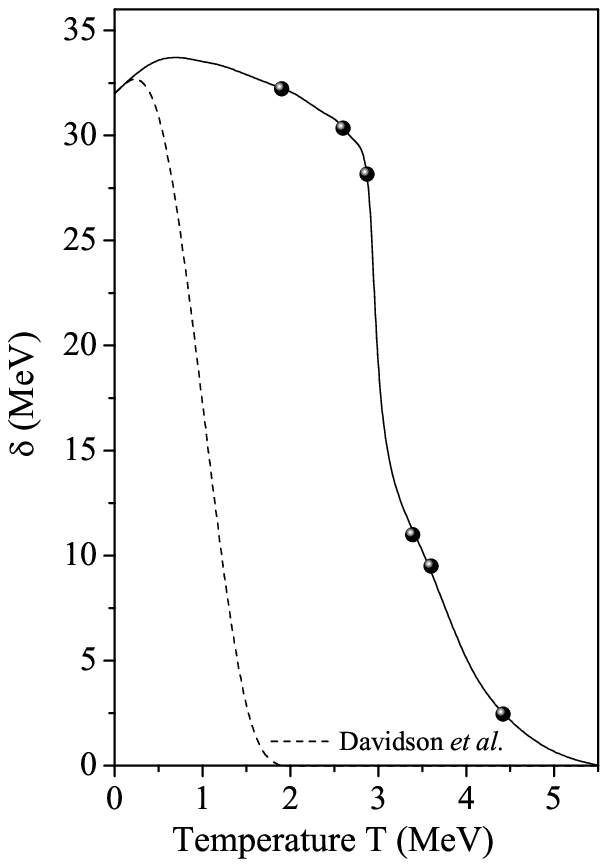}
\vspace*{-0.7cm}
\caption{The pairing energy $\delta$ (MeV) as function of temperature T (MeV), readjusted empirically for temperatures 
T$>$1.5 MeV (solid line and solid dots), compared with the original curve (dashed line) due to Davidson {\it et al.} 
\cite{davidson}.}
\label{fig:4}
\end{center}
\end{figure} 

\section{Calculations and Results}

Table 1 gives the newly fitted constants of Seeger's $V_{LDM}$ for the experimental binding energy $B_{expt}$ 
\cite{audi03}, and the theoretical $B_{theo}$ values \cite{moller95} where the experimental data were not available. 
Interestingly, only the bulk constant $\alpha(0)$, working as an overall scaling factor, and the asymmetry constant $a_a$, 
controlling the curvature of the experimental parabola, are required to be re-adjusted. The role of these re-fitted
constants is illustrated in Fig. \ref{fig:1} for Z=97 nuclides. We notice in Fig. \ref{fig:1} an excellent agreement 
between the present fits (crosses and down open triangles) corresponding to experimental (solid circles) \cite{audi03} and 
theoretical data (open circles) \cite{moller95}, respectively. The fits are with in 0-1.5 MeV of the available $B_{expt}$ 
or $B_{theo}$ data. Also plotted in Fig. \ref{fig:1} are the results of calculations using the old 1961 Seeger's constants
(hollow squares), showing the requirement and extent to which the fitting can clearly improve upon the older results.

Next, we consider an application of the re-adjusted $V_{LDM}$ with an idea to impress upon the need and to propose here 
atleast a partially modified variation of the pairing constant $\delta$ with temperature T, as compared to that of Davidson 
{\it et al.} \cite{davidson}. Fig. \ref{fig:2} shows the fragmentation potential V(A) for the decay of $^{56}$Ni$^{*}$ 
(a complete mass spectrum) into light particles (LPs) and intermediate mass fragments (IMFs) at T=3.60 MeV for two 
different $\ell$ values ($\ell$=0 and 36 $\hbar$), compared with one at T =0 for $\ell$=0 $\hbar$. We notice that at T =0 
for $\ell$=0 $\hbar$, the pairing effects are very strong since all the even-even fragments lie at potential energy minima. 
On the other hand, if we include temperature effects as per prescription of Davidson {\it et al.} (dashed line in 
Fig. \ref{fig:4}), we find that $\delta$=0 MeV in $V_{LDM}$ for T$>$2 MeV, and hence in Fig. \ref{fig:2} for T=3.60 MeV, 
$\delta$=0 MeV, the odd-odd fragments like $^{10}$B, $^{14}$N, $^{18}$F, etc., become equally probable as the even-even 
fragments, since minima are now equally stronger. The same result was obtained earlier in \cite{rkgupta05} for the decay of 
$^{56}$Ni$^{*}$ at T=3.39 MeV, since there too $\delta$=0 MeV was used from Davidson {\it et al.}. However, if we 
empirically choose $\delta$=9.5 MeV for T=3.60 MeV (for the best fit to IMFs data in Fig. \ref{fig:3}), the situation 
becomes again favourable. In other words, Fig. \ref{fig:2} for T=3.60 MeV, $\delta$=9.5 MeV shows once again that the 
even-even fragments, like $^{12}$C, $^{16}$O, etc., are equally favoured as odd-odd $^{14}$N, $^{18}$F, etc. It is 
important to note that in this experiment \cite{sanders89} on $^{32}$S+$^{24}$Mg$\rightarrow ^{56}$Ni$^*$, only the IMFs 
are measured, and theoretically LPs are more prominent at lower $\ell$-values whereas IMFs seem to supersede them at higher 
$\ell$-values, as is also evident from Fig. \ref{fig:2}. The calculated decay cross-sections $\sigma_{IMFs}$ for IMFs at 
T=3.60 MeV, for both $\delta$=0 and 9.5 MeV cases are shown in Fig. \ref{fig:3}, compared with experimental data 
\cite{sanders89}. We notice in this figure that better comparisons are obtained for the case of $\delta\neq$0 calculations, 
contrary to earlier results in Fig. 13 of \cite{rkgupta05} for $\delta$=0 MeV, but supporting the one in Fig. 7 of 
\cite{rkgupta08} for $\delta$=9.5 MeV. Similar calulations, supporting non-zero $\delta$ values at T$>$2 MeV, are also 
reported for $^{56}$Ni$^{*}$ at T=3.39 MeV in Fig. 7 of \cite{rkgupta08}, and for fusion-fission cross-section in 
$^{118}$Ba (Fig. 2(b) in \cite{kumar09}), and the possible $^{14}$C clustering in $^{18,20}$O and $^{22}$Ne nuclei 
\cite{bansal11}. These calculations lead us to modify the variation of $\delta$ as function of T, as shown in 
Fig. \ref{fig:4} (solid line throgh solid dots). Apparently, many more calculations are needed for Fig. \ref{fig:4} to 
represent a true $\delta (T)$.

\begin{table*}
\tabcolsep 0.5mm
\caption{Re-fitted bulk $\alpha$(0) and asymmetry $a_a$ constants for Seeger's liquid drop energy in 1$\leq$Z $\leq$118 
nuclei, {\it w.r.t.} $B_{expt}$ (for nuclei upto Z=7, and Z$\geq$8 marked with star) and $B_{theo}$ (only for Z$\geq$8 
and where experimental data were not available).}
\begin{center}
\begin{small}
\begin{tabular}{|c|c|c|c||c|c|c|c||c|c|c|c||c|c|c|c|}\hline
Z & N & $\alpha (0)$ & a$_a$ & Z & N & $\alpha (0)$ & a$_a$ & Z &N& $\alpha (0)$ & a$_a$ \\\hline
 
1&    2       &-15.85&0.100 &   6&   3,11     &-14.81&0.100 &    10&    25,26               &-16.25&0.856 \\\hline

 &    3       &-16.93&0.120 &    &   4,9      &-15.65&0.100 &      &     27                 &-16.25&0.848\\ \hline
 
 &    4       &-13.37&0.100 &    &5,6$^{\dagger}$,7&-16.53&0.100 & &     28                 &-16.25&0.839 \\ \hline

 &    5       &-13.57&0.120 &    &    8       &-15.92&0.100 &      &     29                 &-16.25&0.833 \\ \hline

 &    6       &-11.49&0.100 &    &   10       &-15.09&0.100 &      &     30                 &-16.25&0.826 \\ \hline

2&    1       &-15.18&0.100 &    &  12,13     &-15.04&0.800 &      &     31                 &-16.25&0.815 \\ \hline

 &    2       &-16.11&0.100 &    &   14       &-14.83&0.800 &    11&    (7,8)$^{*}$         &-15.91&0.920  \\
\hline
  
\end{tabular}
\\ $^{\dagger}$ For Z=6, N=6, $\alpha(0)$=-16.72 instead of -16.53.
\end{small}
\end{center}
\end{table*}

\begin{table*}
\tabcolsep 0.5mm
\begin{center}
\begin{small}
{Continued 1 ....... Table 1}
\begin{tabular}{|c|c|c|c||c|c|c|c||c|c|c|c||c|c|c|c|}\hline
Z & N & $\alpha (0)$ & a$_a$ & Z & N & $\alpha (0)$ & a$_a$ & Z &N& $\alpha (0)$ & a$_a$ \\\hline 

14&    36,37          &-16.09&0.762  &19&    51            &-16.21&0.734 &25&     15,16          &-16.38&0.758 \\ \hline
 
  &    38             &-16.09&0.758  &20&    10            &-16.17&0.834 &  &     17,18          &-16.48&0.780 \\ \hline

  &   39,40           &-16.09&0.754  &  &    11            &-16.17&0.806 &  &    (19-26)$^{*}$   &-16.46&0.760 \\ \hline

15&      8            &-16.10&0.970  &  &    12            &-16.17&0.771 &  &    (27-44)$^{*}$   &-16.42&0.768 \\ \hline

  &     (9-14)$^{*}$  &-16.31&0.939  &  &    13            &-16.17&0.718 &  &     45,46          &-16.43&0.761 \\ \hline

  &    (15-24)$^{*}$  &-16.40&0.893  &  &    (14-21)$^{*}$ &-16.49&0.899 &  &     47-52          &-16.21&0.715 \\ \hline

  &    (25-28)$^{*}$  &-16.12&0.783  &  &    (22-30)$^{*}$ &-16.37&0.789 &  &     53-59          &-16.17&0.706 \\
\hline

\end{tabular}
\end{small}
\end{center}
\end{table*}

\begin{table*}
\tabcolsep 0.5mm
\begin{center}
\begin{small}
{Continued 2 ....... Table 1}\\
\begin{tabular}{|c|c|c|c||c|c|c|c||c|c|c|c||c|c|c|c|}\hline
Z & N & $\alpha (0)$ & a$_a$ & Z & N & $\alpha (0)$ & a$_a$ & Z &N& $\alpha (0)$ & a$_a$ \\\hline
 
32&    23-25          &-16.58&0.720 &38&   (35-39)$^{*}$  &-16.67&0.735 &43&(42,43,70-72)$^{*}$  &-16.73&0.755 \\ \hline

  &   (26-33)$^{*}$   &-16.58&0.749 &  &(40-42,64-67)$^{*}$&-16.63&0.742&  &   (44,45)$^{*}$    &-16.70&0.748  \\ \hline

  &   (34-45)$^{*}$   &-16.55&0.795 &  &   (43-54)$^{*}$  &-16.58&0.748 &  &   (46-54)$^{*}$    &-16.66&0.787  \\ \hline

  &   (46-57)$^{*}$   &-16.41&0.719 &  &   (55-63)$^{*}$  &-16.57&0.734 &  &   (55-69)$^{*}$    &-16.61&0.728  \\ \hline

  &     58            &-16.34&0.701 &  &    68-71         &-16.47&0.705 &  &   (73-75)$^{*}$    &-16.72&0.748  \\ \hline

  &    59-80          &-16.32&0.701 &  &    72-79         &-16.47&0.708 &  &    76-80           &-16.55&0.711  \\ \hline

33&    24-26          &-16.60&0.720 &  &    80-85         &-16.21&0.673 &  &    81-88           &-16.55&0.707  \\
\hline

\end{tabular}
\end{small}
\end{center}
\end{table*}

\begin{table*}
\tabcolsep 0.5mm
\begin{center}
\begin{small}
{Continued 3 ....... Table 1}\\
\begin{tabular}{|c|c|c|c||c|c|c|c||c|c|c|c||c|c|c|c|}\hline
Z & N & $\alpha (0)$ & a$_a$ & Z & N & $\alpha (0)$ & a$_a$ & Z &N& $\alpha (0)$ & a$_a$ \\\hline 

47&    84-93         &-16.59&0.705 &52&    47-50         &-16.86&0.680 &56&   (77-92)$^{*}$  &-16.73&0.712  \\ \hline

  &    94-95         &-16.57&0.701 &  &    51,52         &-16.84&0.630 &  &    98-120        &-16.70&0.700  \\ \hline

  &    96-101        &-16.39&0.677 &  &   (53-59)$^{*}$  &-16.81&0.885 &  &    121-131       &-16.54&0.681  \\ \hline

  &    102-105       &-16.34&0.672 &  &   (60-66)$^{*}$  &-16.78&0.796 &  &    132,133       &-16.54&0.680  \\ \hline

  &    106-113       &-16.39&0.679 &  &   (67-79)$^{*}$  &-16.70&0.723 &57&    53-56         &-16.91&0.600  \\ \hline

48&   42-44          &-16.80&0.678 &  &   (80-90)$^{*}$  &-16.69&0.713 &  &    57,58         &-16.89&0.600  \\ \hline

  &   45,46          &-16.78&0.695 &  &    91-98         &-16.64&0.702 &  &    59            &-16.88&0.600  \\
\hline

\end{tabular}
\end{small}
\end{center}
\end{table*}

\begin{table*}
\tabcolsep 0.5mm
\begin{center}
\begin{small}
{Continued 4 ....... Table 1}\\
\begin{tabular}{|c|c|c|c||c|c|c|c||c|c|c|c||c|c|c|c|}\hline
Z & N & $\alpha (0)$ & a$_a$ & Z & N & $\alpha (0)$ & a$_a$ & Z &N& $\alpha (0)$ & a$_a$ \\\hline 

61&   (78-85)$^{*}$   &-16.84&0.745 &66&    67,68          &-17.02&0.826 &70&   124-131        &-16.86&0.705 \\ \hline

  &   (86-102)$^{*}$  &-16.75&0.695 &  &    69,70          &-17.01&0.826 &  &   132-164        &-16.65&0.672 \\ \hline

  &    103-105        &-16.79&0.705 &  &    71             &-16.99&0.826 &71&   75             &-17.03&0.603 \\ \hline

  &    106-118        &-16.79&0.707 &  &   (72-77)$^{*}$   &-17.00&0.930 &  &   76-78          &-17.03&0.910 \\ \hline

  &    119-128        &-16.62&0.684 &  &   (78-85)$^{*}$   &-16.92&0.768 &  &  (79-88)$^{*}$   &-17.00&0.797 \\ \hline

  &    129-144        &-16.43&0.662 &  &   (86-91)$^{*}$   &-16.84&0.700 &  &  (89-113)$^{*}$  &-16.93&0.717 \\ \hline

62&    61,62          &-17.00&0.600 &  &   (92-107)$^{*}$  &-16.86&0.709 &  &   114-124        &-16.95&0.723 \\
\hline

\end{tabular}
\end{small}
\end{center}
\end{table*}

\begin{table*}
\tabcolsep 0.5mm  
\begin{center}
\begin{small}
{Continued 5 ........ Table 1}
\begin{tabular}{|c|c|c|c||c|c|c|c||c|c|c|c||c|c|c|c|}\hline
Z & N & $\alpha (0)$ & a$_a$ & Z & N & $\alpha (0)$ & a$_a$ & Z &N& $\alpha (0)$ & a$_a$ \\\hline 
 
76&    137-142       &-16.72&0.671 &82&    134-143        &-17.00&0.712&87&   111            &-17.25&0.845 \\ \hline 
                 
  &    143-152       &-16.72&0.672 &  &    144-159        &-16.86&0.685&  & (112-117)$^{*}$  &-17.14&0.759 \\ \hline

  &    153-163       &-16.73&0.674 &  &    160-165        &-16.87&0.687&  & (118-129)$^{*}$  &-17.09&0.728 \\ \hline

  &    164-166       &-17.02&0.705 &  &    166-188        &-16.86&0.685&  & (130-145)$^{*}$  &-17.01&0.700 \\ \hline

  &    167-170       &-17.03&0.705 &  &    189-191        &-16.86&0.684&  &   146-153        &-16.95&0.688 \\ \hline

  &    171-177       &-16.80&0.683 &83&    95,96          &-17.14&0.845&  &   154-163        &-16.95&0.690 \\ \hline

77&    85,86         &-17.09&0.845 &  &    97-100         &-17.15&0.845&  &   164-189        &-16.95&0.691 \\
\hline
 
\end{tabular}
\end{small}
\end{center}
\end{table*}

\begin{table*}
\tabcolsep 0.5mm
\begin{center}
\begin{small}
{Continued 6 ....... Table 1}\\
\begin{tabular}{|c|c|c|c||c|c|c|c||c|c|c|c||c|c|c|c|}\hline
Z & N & $\alpha (0)$ & a$_a$ & Z & N & $\alpha (0)$ & a$_a$ & Z &N& $\alpha (0)$ & a$_a$ \\\hline 

90&   155-171        &-17.06&0.703 &93&    152-155       &-17.09&0.701 &96&   132,133        &-17.34&0.803 \\ \hline

  &   177-181        &-17.07&0.703 &  &    156-186       &-17.10&0.705 &  &   134,135        &-17.36&0.803 \\ \hline

  &   182-189        &-16.92&0.685 &  &    187-189       &-17.00&0.691 &  &   136            &-17.37&0.800 \\ \hline

  &   190-192        &-16.92&0.684 &  &    190-192       &-17.00&0.690 &  &(137,154-156)$^{*}$&-17.15&0.707\\ \hline

  &   193-196        &-16.92&0.683 &  &    193-195       &-17.00&0.689 &  & (138-153)$^{*}$  &-17.15&0.705 \\ \hline

  &   197-202        &-16.92&0.682 &  &    196-199       &-17.00&0.688 &  &   157-186        &-17.15&0.709 \\ \hline

  &   203-205        &-16.92&0.681 &  &    200-203       &-17.00&0.687 &  &   187-189        &-17.02&0.691 \\
\hline

\end{tabular}
\end{small}
\end{center}
\end{table*}

\begin{table*}
\tabcolsep 0.5mm
\begin{center}
\begin{small}
{Continued 7 ....... Table 1}
\begin{tabular}{|c|c|c|c||c|c|c|c||c|c|c|c||c|c|c|c|}\hline

Z & N & $\alpha (0)$ & a$_a$ & Z & N & $\alpha (0)$ & a$_a$ & Z &N& $\alpha (0)$ & a$_a$ \\\hline

98&    196-198       &-17.05&0.689 &101&    134,135      &-17.37&0.801 &103&   171-188        &-17.24&0.717 \\ \hline

  &    199-201       &-17.05&0.688 &   &    136,137      &-17.39&0.803 &   &   189-190        &-17.08&0.690 \\ \hline

  &    202-205       &-17.04&0.686 &   &    138,139      &-17.40&0.801 &   &   191-193        &-17.08&0.689 \\ \hline

  &    206-211       &-17.04&0.685 &   &    140,141      &-17.42&0.802 &   &   194-196        &-17.08&0.688 \\ \hline

  &    212-227       &-17.04&0.684 &   &    142,143      &-17.44&0.803 &   &   197-199        &-17.08&0.687 \\ \hline

99&    125-129       &-17.31&0.800 &   &  (144-155)$^{*}$&-17.22&0.710 &   &   200-205        &-17.08&0.686 \\ \hline

  &    130,131       &-17.33&0.800 &   &  (156-161)$^{*}$&-17.22&0.712 &   &  206-215,227     &-17.08&0.685 \\
\hline

\end{tabular}
\end{small}
\end{center}
\end{table*}

\begin{table*}
\tabcolsep 0.5mm
\begin{center}
\begin{small}
{Continued 8 ....... Table 1}
\begin{tabular}{|c|c|c|c||c|c|c|c||c|c|c|c||c|c|c|c|}\hline
Z & N & $\alpha (0)$ & a$_a$ & Z & N & $\alpha (0)$ & a$_a$ & Z &N& $\alpha (0)$ & a$_a$ \\\hline

106&    144,145       &-17.46&0.802 &109&    150,151          &-17.51&0.803 &112&   214-224         &-17.17&0.687 \\ \hline

   &    146,147       &-17.47&0.800 &   &    152,153          &-17.52&0.803 &   &   225-227         &-17.17&0.686 \\ \hline

   &    148,149       &-17.49&0.802 &   &    154              &-17.53&0.803 &113&   153-164         &-17.38&0.728 \\ \hline

   &    150,151       &-17.50&0.800 &   &    155              &-17.34&0.724 &   &   165-169         &-17.37&0.727 \\ \hline

   &  (152-165)$^{*}$ &-17.33&0.731 &   &   (156-162)$^{*}$   &-17.34&0.727 &   &  (170-174)$^{*}$  &-17.35&0.724 \\ \hline

   &  (166-167)$^{*}$ &-17.33&0.731 &   &   (163-170)$^{*}$   &-17.33&0.726 &   &   175-188         &-17.37&0.729 \\ \hline

   &    168-186       &-17.29&0.722 &   &    171-183          &-17.33&0.726 &   &   189-197         &-17.18&0.690 \\
\hline

\end{tabular}
\end{small}
\end{center}
\end{table*}

\section{Summary}

In view of the large data for ground state (g.s.) binding energies having become available and to be able to include the 
T-dependence on binding energies, we have re-fitted two of the constants, the bulk $\alpha(0)$ and neutron-proton asymmetry 
$a_a$, of Seeger's mass formula.  The experimental g.s. binding energies or theoretical binding energies for neutron- and 
proton-rich nuclei, where data are not yet available, are fitted within $<$1.5 MeV, and up to Z=118 nuclei. The method used
is the Strutinsky renormalization procedure to define the g.s. binding energy as a sum of the liquid drop energy and the 
shell correction. Taking shell correction from the empirical formula of Myers and Swiatecki, the two constants of Seeger's 
liquid drop energy are fitted to obtain the experimental or theoretical binding energy. The fitted constants of liquid drop 
energy have been used for understanding the dynamics of excited compound nuclear systems, which point out to the 
inadequacy of the variation of pairing energy constant $\delta$ with temperature T. As per the given $\delta(T)$ variation 
of Davidson {\it et al.}, $\delta$=0 MeV for T$>$2 MeV. However, the recent compound nucleus decay calculations suggest
that $\delta\neq$0 for T$>$ 2 MeV and hence clearly indicate the need for re-evaluation of the T-dependence of Seeger's 
constants. A new dependence of $\delta(T)$ is suggested on the basis of already published calculations for compound nucleus
decay studies. Need for further studies are clealy indicated.

\acknowledgments
The authors are thankful to Prof. S. K. Patra for his interest in this work. M.K.S. is thankful to CSIR, New Delhi, and
R.K.G. to Department of Science and Technology, Govt. of India, for financial support for this work in the form of 
research projects.

\end{document}